\documentclass[iop]{emulateapj}
\usepackage{amsmath}
\usepackage{amssymb}
\usepackage{graphicx}
\usepackage{epstopdf}
\usepackage{natbib}
\usepackage{color}

\begin{document}

\title{BL LAC PKS\,B1144$-$379 an extreme scintillator}
\author{R. J. Turner, S. P. Ellingsen, S. S. Shabala, J. Blanchard, J. E. J. Lovell, and J. N. McCallum}
\affil{School of Mathematics and Physics, Private Bag 37, University of Tasmania, Hobart, TAS 7001, Australia}
\author{G. Cim\`{o}}
\affil{Joint Institute for VLBI in Europe, Postbus 2, 7990 AA Dwingeloo, Netherland}

\begin{abstract}
Rapid variability in the radio flux density of the BL Lac object PKS\,B1144$-$379 has been observed at four frequencies, ranging from 1.5 to 15~GHz, with the VLA and the University of Tasmania's Ceduna antenna. Intrinsic and line of sight effects were examined as possible causes of this variability, with interstellar scintillation best explaining the frequency dependence of the variability timescales and modulation indices. This scintillation is consistent with a compact source 20--40$~{\mu}$as, or 0.15--0.3~pc in size. The inferred brightness temperature for PKS\,B1144$-$379 (assuming that the observed variations are due to scintillation) is $6.2 \times 10^{12}$~K at 4.9~GHz, with approximately 10 percent of the total flux in the scintillating component. We show that scintillation surveys aimed at identifying variability timescales of days to weeks are an effective way to identify the AGN with the highest brightness temperatures.

\end{abstract}

\keywords{galaxies: active---scattering---ISM: structure}

\section{Introduction}
The BL Lac object PKS\,B1144$-$379 \citep{Nicolson et al.+1979} ($11^{\rm h} 47^{\rm m} 01^{\rm s} .4$, $-38^{\circ} 12^{\prime} 11^{\prime\prime}$ in J2000 coordinates), located at $z=1.048$ \citep{Stickel et al.+1989} is known for both its long and short-term flux density variability at centimetre wavelengths. Some of the first observations of PKS\,B1144$-$379 showed variability at a frequency of 5 GHz. The flux density increased from 0.9 Jy in 1970 December to 1.6 Jy in 1971 February and again to 2.22 Jy in 1971 September \citep{Shimmins Bolton+1972,Bolton Shimmins+1973,Gardner et al.+1975}. Between May and August 1994, the flux density of PKS\,B1144$-$379 at 4.8 GHz dropped 17\%, and 9\% at 8.6 GHz \citep{Kedziora-Chudczer et al.+2001b}.

Long-term (months to years) variability is an indicator of intrinsic variations in the source. For example, VLBI imaging of PKS\,B1144$-$379 shows changes in the extended jet-like components on milliarcsecond scales over these timescales, in addition to variations in the flux density of the quasar core \citep{Ojha et al.+2010}. This potentially degrades the utility of this object as an international celestial reference frame (ICRF) defining source \citep{Ma et al.+2009}.

Intraday variability in PKS\,B1144$-$379 was first identified by \citet{Kedziora-Chudczer et al.+2001a} and on this basis it was included in the long term 6.7 GHz monitoring which commenced 2003 April 3 using the University of Tasmania's 30 m Ceduna antenna as part of the Continuous Single-dish Monitoring of Intraday variability at Ceduna (COSMIC) project \citep{McCulloch et al.+2005,Carter et al.+2009}. Observations between 2003 and 2010 measured the mean 6.7 GHz flux density to range from 0.8 to 3.2 Jy, with the largest amplitude variations occurring on timescales of hundreds of days (the data from this longer term monitoring will be presented as part of a future publication).

Quasars are known to exhibit short timescale variability and in particular, intraday variability \citep{Wagner Witzel+1995}. Sources PKS\,B0405$-$385, PKS\,B1257$-$326, and J1819+3845 \citep{Kedziora-Chudczer et al.+1997,Bignall et al.+2006,Dennett-Thorpe de Bruyn+2000} have been observed to vary on timescales of less than 2 hours at centimeter wavelengths. These variations can either be due to intrinsic or line of sight effects. The assumption of rapid intrinsic variability typically requires source brightness temperatures well in excess of the inverse Compton limit ($\sim 10^{12}$ K), or unrealistically high Doppler beaming factors \citep[see for example][]{Kedziora-Chudczer et al.+1997}.

Extrinsic effects include jet precession \citep[e.g.][]{Camenzind Krockenberger+1992}, gravitational microlensing \citep[e.g.][]{Paczynski+1986} and interstellar scintillation (ISS) \citep[e.g.][]{Walker+1998}. Of these, ISS best fits the observed frequency dependence, annual cycles and time-delay for the variations observed in objects such as PKS\,B0405$-$385, PKS\,B1257$-$326, and J1819+3845.  Limited studies of the short timescale variability of PKS\,B1144$-$379 exist. In 1999 May, \citet{Kedziora-Chudczer et al.+2001a} observed PKS\,B1144$-$379 at four frequencies over a period of six days.  Intra day variability (IDV) in the flux density on timescales of 3 to 4 days was observed and \citeauthor{Kedziora-Chudczer et al.+2001a} suggested interstellar scintillation as the likely cause. The MASIV survey of 443 compact radio sources demonstrated that where variations on timescales of days are observed in AGN at centimetre wavelengths, scintillation is the predominant cause \citep{Lovell et al.+2003,Lovell et al.+2008}.

Interstellar scintillation is the interference phenomenon seen when radiation from a distant source passes through the turbulent, ionised interstellar medium (ISM) of our Galaxy. Interference arises due to small variations in the medium's refractive index for different paths through the ISM. As this interstellar material moves with respect to the observer, intensity variations will be observed as a function of time. Scintillation behaviour is characterised as either strong or weak scattering. In weak scattering only small phase changes are introduced by the ISM over the first Fresnel zone. In strong scattering the wavefront is rapidly varying on scales smaller than the first Fresnel zone. Two types of variability are expected in the strong scattering regime, slow broadband variability due to refractive scattering and fast narrowband variability due to diffractive scattering. Unlike jet precession or microlensing, the measured ISS variability timescales are expected to be frequency dependent \citep{Walker+1998}. 

Observations of PKS\,B1144$-$379 during 2004 with the Ceduna antenna at 6.7 GHz showed both the flux density and the amplitude of variability to be decreasing.  However, lower amplitude quasi-periodic variations with a period in excess of a week persisted throughout this ``quiet'' phase and the purpose of the observations presented here is to use multi-frequency radio data to determine the mechanism for these variations.  In this paper, we report 15 days of observations of PKS\,B1144$-$379 at frequencies of 1.5, 4.9 and 15 GHz using the Very Large Array (VLA), and 6.7 GHz observations using the University of Tasmania's Ceduna antenna.

\section{Observations and data reduction}
\label{sec:obs}
The source PKS\,B1144$-$379 was observed over a 15 day period between 2005 January 10 and 2005 January 24 using the VLA. Observations were made at three frequencies centered at 1.5, 4.9 and 15 GHz, each with a 50 MHz bandwidth. Primary flux density calibration was obtained through observations of 1331+305 (3C286) for each of the frequency bands. Daily observations were made for each frequency for a duration of approximately 3 minutes for 1331+305 and approximately 10 minutes for PKS\,B1144$-$379. These observations were conducted during the reconfiguration of the VLA (from the A array to the B array), and consequently on some days not all antennas were available (typically no more than 2 or 3 antennas were missing from the array).

The data collected using the VLA was processed using the Astronomical Image Processing System (AIPS)\footnote{http://www.aips.nrao.edu/index.shtml}. Standard VLA calibration and data reduction procedures were used for each frequency and each day, with only minor differences (e.g. changes to the averaging time for self-calibration). Antennas or time ranges for which the data were clearly erroneous were removed prior to calibration. The data for PKS\,B1144$-$379 were calibrated against 1331+305 for which the assumed flux densities were 1.16, 0.87 and 0.54 Jy at the frequencies 1.5, 4.9 and 15 GHz respectively \citep{Baars et al.+1977}. The data for PKS\,B1144$-$379 were then self-calibrated using an initial model of a point source of peak intensity 1 Jy. The resulting image was cleaned with clean boxes placed around PKS\,B1144$-$379 and other visible sources within the primary beam. This image was self-calibrated using only phases, then cleaned, with this procedure repeated a further time correcting both the amplitudes and phases. The flux densities of all sources were measured from the final image using the AIPS task SAD. Typical noise levels in the final images were of the order of 1 mJy, and this is taken as representative of the uncertainty in our flux density measurements.

The University of Tasmania's Ceduna antenna in South Australia was also used to observe PKS\,B1144$-$379 over a longer period as part of the COSMIC project \citep{McCulloch et al.+2005}, with observations commencing 2003 April 3 and still being undertaken at the time of writing. Observations of PKS\,B1144$-$379 were made on just over half the days in this time range in contiguous blocks, typically ranging from 5 to 15 days. The observing frequency band has a center frequency of 6.7 GHz and a 500 MHz bandwidth for each of two orthogonal circular polarizations. Around 50 independent flux density measurements of PKS\,B1144$-$379 are made each day as part of COSMIC \citep[see][]{Carter et al.+2009}.  To improve the signal to noise ratio here we present the daily mean flux density measured with the Ceduna antenna. Data from the Ceduna antenna were only available for 11 of the 15 days for which observations were made with the VLA.

\section{Results}
\begin{figure}
\begin{center}
\includegraphics[width=8cm]{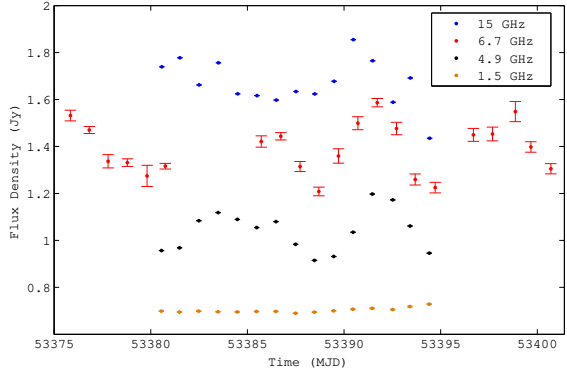}
\end{center}
\caption{The 1.5, 4.9 and 15 GHz light curves (orange, black and blue respectively) of PKS\,B1144$-$379 observed with the VLA and the 6.7 GHz light curve (red) observed with the University of Tasmania Ceduna antenna over the period MJD 53375 -- 53400 (2005 January 5--30).}
\label{fig:lightcurve}
\end{figure}

The light curves of PKS\,B1144$-$379 observed with both the VLA and the Ceduna antenna are plotted in Figure ~\ref{fig:lightcurve}.   To assess the variability timescale we used the approach of \citet{Lovell et al.+2008} to produce structure functions for each frequency. The characteristic timescale is defined to be the time for the structure function to reach half the value of its maximum \citep{Lovell et al.+2008}. At both 4.9 and 6.7 GHz, a characteristic timescale of between 1 and 1.5 days was estimated. At 15 GHz there is evidence of two timescales of approximately 0.4 and 2 days. A complete cycle was not observed at 1.5 GHz in the 15 day sampling interval, so conservatively, the characteristic timescale is greater than 4.8 days. The peak-to-peak variability timescales evident in Figure~\ref{fig:lightcurve} are $2\pi$ times greater than these characteristic timescales. Due to the low number of data points, we sought to validate these timescale estimates by crudely fitting sinusoids to the data points. The estimated peak-to-peak timescale for the 4.9 GHz frequency is 7.7 days, and for the 6.7 GHz frequency it is 6.6 days. Ordinary least-squares regression analysis\footnote{In ordinary least-squares analysis, $R^2$ value represents the proportion of variability in the data that can be accounted for by a given model. Values of $R^2$ close to one thus represent a good fit, in our case suggesting that the observed light curves can be approximated by sinusoidal functions.} yields $R^2$ values of 0.76 and 0.83, respectively, for these fits. For the 15 GHz frequency, least-squares analysis shows a timescale of 8.6 days ($R^2=0.57$), and a weaker fit ($R^2=0.26$) at 1.7 days. In Figure~\ref{fig:timescales} we plot the resultant characteristic timescales. These estimates are consistent with the characteristic timescales estimated from the structure functions. Peak-to-peak timescales of approximately 2 days are below the Nyquist frequency with the daily sampling regime. Observations of PKS\,B1144$-$379 taken in 1999 by \citet{Kedziora-Chudczer et al.+2001a} clearly show a timescale at 8.6 GHz 3 -- 4 times shorter than at 4.8 GHz. However, the source is likely to have undergone significant intrinsic evolution since then, as evidenced by the 4.8 GHz flux falling from 2\,Jy mean in May 1999 to 1\,Jy in January 2005.

The Discrete Correlation Function \citep{Edelson Krolik+1988} was used to determine if the light curves of any of the observed frequencies are correlated. The correlation between the 4.9 and 6.7 GHz observations is greatest ($\sim 1$) for a time lag of between 0.5 and 1 days (4.9 GHz lagging 6.7 GHz). None of the other frequency pairs showed a correlation coefficient above 0.6.

\begin{figure}
\begin{center}
\includegraphics[width=8cm]{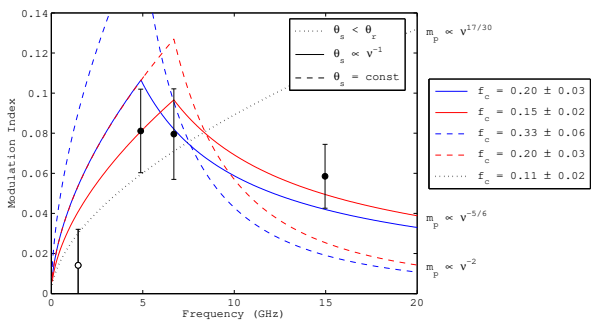} 
\end{center}
\caption{The modulation index as a function of frequency. The uncertainty in the modulation index $m$ is estimated as $\sigma_{\rm m}=0.9\,m(t_{\rm char}/{\rm 15\,days})^{1/2}$, where 15 days is the observing interval \citep{Stinebring et al.+2000}. Also plotted is a one-parameter ($f_{\rm c}$) theoretical fit for the expected scintillation behaviour. The rising part of each curve corresponds to point-like scintillation. The falling part of each curve corresponds to quenched scintillation. The coloured lines represent models where source size exceeds the refractive scale at 4.9 (blue) and 6.7 (red) GHz. Solid lines are for models where source size $\propto \nu^{-1}$. Dashed lines are models where source size is independent of frequency. Unquenched scintillation at all frequencies (dotted line) clearly cannot explain the data.}
\label{fig:modind}
\end{figure}

\begin{figure}
\begin{center}
\includegraphics[width=8cm]{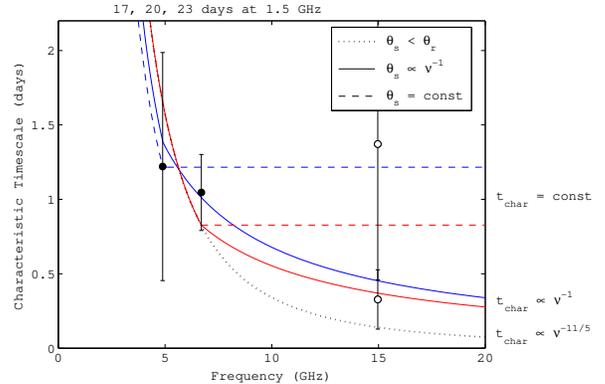} 
\end{center}
\caption{Variability timescales as a function of frequency. Only lower limits are available at 1.5 GHz; these are shown numerically at the top of each curve. 15 GHz variability suggests two contributing timescales. The smaller formal uncertainty at 6.7 GHz is due a longer time series for Ceduna compared with the VLA data. The break in model curves corresponds to a frequency above which source size exceeds the refractive scale. Colours and line styles are as in Figure~\ref{fig:modind}.}
\label{fig:timescales}
\end{figure}

The modulation index, $m$, was calculated as $m=\sigma/\langle S \rangle$ with $\sigma$ the standard deviation of the flux density and $\langle S \rangle$ the mean flux density. The modulation indices calculated from the observations for the 1.5, 4.9, 6.7 and 15 GHz frequencies are 0.014, 0.081, 0.080 and 0.059 respectively. These are plotted in Figure~\ref{fig:modind}. Since only a fraction of a cycle is sampled over the 15 day observation period at 1.5 GHz, the derived modulation index is a lower limit. Comparisons with the MASIV sample \citep{Lovell et al.+2008} show that the high 4.9 GHz modulation index of PKS\,B1144-379 places it in the top 1--2 \% (in terms of fractional variability) of bright ($>1$~Jy) quasars.

At 1.5 and 4.9 GHz, additional sources are visible within the primary beam of the VLA for the observations of PKS\,B1144$-$379. At both these frequencies, a source was observed with declination and right ascension offset from PKS\,B1144$-$379 by 107 and -99 arcseconds on the sky respectively, while a second source was observed at only 1.5 GHz offset in right ascension and declination by -201 and 189 arcseconds on the sky respectively.    These sources were found to have a constant flux density (within the measurement uncertainty) over the 15 day period of our observations.  This, combined with the close agreement between the pattern of the Ceduna 6.7 and VLA 4.9 GHz variations demonstrates that the observed variability has not been produced by errors in the calibration or data reduction process. Figure~2 of \citet{Carter et al.+2009} shows that there are systematic variations in the Ceduna flux densities of around 0.15~Jy on timescales less than a day.  However, the close agreement between the pattern of variations at 4.9 GHz from the VLA and 6.7 GHz from Ceduna demonstrates that the daily average flux densities produced from the Ceduna 30~m (which have been used here) are very reliable.

\section{Discussion}
The observed variability timescales of approximately 7 days can potentially be explained by either line of sight effects or intrinsic mechanisms. The former includes interstellar scintillation, jet precession or microlensing, and the latter rapid intrinsic variability.

\subsection{Brightness temperature due to intrinsic variability}

If the observed variability in the source is intrinsic, then using causality to limit the linear scale we can infer the brightness temperature of the source. Using Planck's law we can determine the brightness temperature $T_{\rm B}$ (in K) from the variation in flux density $\triangle S$ (in Jy), the frequency $\nu$ (in GHz), the peak-to-peak variability timescale $\tau$ (in days) and the angular diameter distance $D_{\rm A}$ (in Gpc)

\begin{equation}
T_{\rm B}=5.9 \times 10^{15}\left[\frac{\triangle S}{\mathrm{1\:Jy}}\right]\left[\frac{\nu}{\mathrm{GHz}}\right]^{-2}\left[\frac{\tau}{\mathrm{100\:days}}\right]^{-2}\left[\frac{D_{\rm A}}{\mathrm{1\:Gpc}}\right]^2\   \mbox{K}
\label{brightness}
\end{equation}

For our 4.9 GHz VLA observations, the variation in flux density is $\triangle S \sim \pm 0.14$ Jy and the variability timescale is $\tau \sim 7.7$ days. PKS\,B1144$-$379 has a redshift of $z=1.048$ and an angular diameter distance of $D_{\rm A} = 1.677$ Gpc (using standard $\Lambda$CDM with $H_0$ = 71\,km\,s$^{-1}$\,Mpc$^{-1}$ ; $\Omega_m = 0.27$ ; $\Omega_\Lambda = 0.73$). Thus the implied brightness temperature is $T_{\rm B}=1.7 \times 10^{16}$~K. When mapped into the rest frame of the source the brightness temperature is reduced by a factor of $\mathcal{D}^3(1+z)^{-3}$, under the hypothesis of intrinsic variation. At high brightness temperatures, $T_{\rm B} \geqslant 10^{12}$ K, energy losses of radiating electrons due to inverse Compton scattering become extremely large, resulting in a rapid cooling of the system and thereby bringing the brightness temperature quickly below this value \citep{Kellermann Pauliny-Toth+1969,Readhead+1994}. So assuming a source brightness temperature of $T_{\rm B} \leqslant 10^{12}$ K, a Doppler factor of $\mathcal{D} > 53$ is required.  This is a large Doppler factor. Although it is in principle possible for the observed variations to be intrinsic, as we show in \S~\ref{sec:scintill}, the frequency dependence does not favor an intrinsic interpretation.


\subsection{Precession \& microlensing}

Quasi-periodic variation in the flux density can be caused by jet precession close to the line of sight. Using the model of \citet{Camenzind Krockenberger+1992} we can replicate the characteristics of light curves observed at 4.9 GHz for PKS\,B1144$-$379, namely the 7.7 day period and 0.081 modulation index, with realistic parameters for beaming angle, black hole mass and jet opening angle. However, just like intrinsic variability, precession is expected to produce light curves whose period is independent of frequency. This is clearly in disagreement with our data (Figure~\ref{fig:lightcurve}). Thus, jet precession does not provide a viable explanation for the observed variability.  Microlensing is ruled out on two counts, since it is expected to be both achromatic and not quasi-periodic.

\subsection{Scintillation} \label{sec:scintill}
The Galactic coordinates of PKS\,B1144$-$379 are 289.24 degrees longitude and 22.95 degrees latitude. For this line of sight, the transition frequency estimated by the \citet{Cordes Lazio+2001} model of the Galactic electron distribution is $\nu_0 \sim 14.4$ GHz. Thus, all our observing frequencies are expected to be in the refractive scattering regime ($\nu < \nu_0$), with the exception of 15\,GHz which is close to the transition frequency. We can use our multi-frequency timescales and modulation indices to infer physical characteristics of the source.

The refractive angular scale is given by
\begin{equation}
	\theta_{\rm r} = \theta_{\rm r0} (\nu/\nu_0)^{-11/5}
\label{eqn:refScale}
\end{equation}
where $\theta_{\rm r0} = 2.3$~$\mu$as at the Galactic coordinates of interest \citep{Walker+2001}. For a point source, the frequency dependence of the scintillation timescale is thus $t_{\rm char} \propto \nu^{-11/5}$. If the source size evolves with frequency slower than $\nu^{-11/5}$, it is possible that at some $\nu > \nu_{\rm crit}$ the source size exceeds the refractive scale. At that point, the scintillation is quenched, and the timescale will take on the same frequency dependence as source size. There are a number of possible models for source size evolution with frequency. Here we consider the limiting cases $\theta_{\rm s} \propto \nu^{-1}$ (corresponding to a self-absorbed synchrotron source), and $\theta_{\rm s}=$const. While in principle the flux density of the compact component (and, hence, the scintillating fraction of total flux) may change with frequency, we refrain from introducing this extra, highly uncertain free parameter into our models.

Comparison of the 4.9 and 6.7 GHz timescales (Figure~\ref{fig:timescales}) places an upper limit on source size. These characteristic timescales are \citep{Walker+1998}
\begin{equation}
t_{\rm char} = %
\begin{cases}
t_{\rm r0}(\nu_0/\nu)^{11/5} & \text{if } \theta_{\rm s} < \theta_{\rm r}\\
t_{\rm r0}(\theta_{\rm s}/\theta_{\rm r0}) = \theta_{\rm s} \left( D_{\rm screen} / V_{\rm screen} \right) & \text{if } \theta_{\rm s} \geq \theta_{\rm r} .
\end{cases}
\label{time}
\end{equation}
where $t_{\rm r0}$ is the light crossing time of the refractive scale at the transition frequency $\nu_0=14.4$~GHz; and $D_{\rm screen}$ and $V_{\rm screen}$ are the distance and transverse velocity of the screen causing the scintillation.

From Equation~\ref{time}, if the source still appears point-like at 6.7~GHz (i.e. if $\theta_{\rm s}(6.7)<\theta_{\rm r}(6.7)$), we expect the 4.9~GHz scintillation timescale to be longer than the 6.7~GHz timescale by a factor of $(6.7/4.9)^{11/5}=2.0$. This does not appear to be consistent with the data (Figure~\ref{fig:lightcurve}). Formal uncertainties on the characteristic timescales are given by $\sigma_{\rm t\,char}=2.2\,t_{\rm char}(t_{\rm char}/{\rm 15\,days})^{1/2}$, where 15 days is the observing interval \citep{Stinebring et al.+2000}. This yields quite large formal uncertainties of $0.7$ and $0.6$~days for the 4.9 and 6.7~GHz timescales, respectively. This is largely due to not sampling enough scintles over the 15-day observing period. Using the available 6.7\,GHz Ceduna data for an additional 70 days, the same $t_{\rm char}$ is recovered, but with a much smaller uncertainty of 0.2~days. Furthermore, as reported in Section~\ref{sec:obs} we find a strong correlation between the 6.7 and 4.9 GHz light curves over the 15-day observing period. This makes our timescales highly inconsistent with a $\nu^{-11/5}$ evolution over the range 4.9 -- 6.7~GHz. Thus, $\theta_{\rm s}(6.7) \geq \theta_{\rm r}(6.7)$.

The frequency evolution of the modulation index (Figure~\ref{fig:modind}) provides a further constraint. This is given by \citep{Walker+1998}
\begin{equation}
	m_{\rm p} = %
	\begin{cases}
	f_{\rm c} (\nu/\nu_0)^{17/30} & \text{if } \theta_{\rm s} < \theta_{\rm r}\\
	f_{\rm c} (\nu/\nu_0)^{-2} \left( \theta_{\rm s} / \theta_{\rm r0} \right)^{-7/6} & \text{if } \theta_{\rm s} \geq \theta_{\rm r} .
\end{cases}
\label{eqn:modIndex}
\end{equation}
where $f_{\rm c}$ is the fraction of total flux that scintillates; this may in general be a function of frequency.

The observed modulation index (Figure~\ref{fig:modind}) decreases between 6.7 and 15 GHz. This is clearly inconsistent with a $\nu^{17/30}$ evolution expected for a point source, and provides further confirmation that we begin to see source structure rather than the refractive scale at some $\nu_{\rm crit}<6.7$\,GHz. By fitting a series ($\nu_{\rm crit}=$ 4.9, 6.7, 20\,GHz) of simple one-parameter curves (for scintillating fraction $f_{\rm c}$), we find that models in which source size is independent of frequency consistently underpredict the modulation index at 15 GHz. The $\theta_{\rm s} \propto \nu^{-1}$ model provides a better fit to the data, requiring source size to exceed the refractive scale at some $\nu_{\rm crit} > 4.9$~GHz.

A final consistency check comes from our lower limit on the 1.5\,GHz timescale, $t_{\rm char}>4.8$~days. Fitting a transition from $\nu^{-11/5}$ to $\nu^{-1}$ dependence at $\nu=4.9$\,GHz through our timescale data yields a 1.5~GHz timescale of 20 days, well in excess of the expected lower limit of 4.8 days.

Knowledge of the transition frequency at which scintillation becomes quenched allows us to estimate the size of the scintillating core. Equation~\ref{eqn:refScale} gives the refractive scales at 4.9 and 6.7~GHz as 25 and 12\,$\mu$as respectively. The discussion above suggests the scintillation becomes quenched in this frequency range, and we therefore estimate the source size (Gaussian FWHM) as between 20 and 41\,$\mu$as at these frequencies\footnote{A factor of $2 \sqrt{\ln 2}$ between refractive scale and source size comes about because observationally the source size is typically defined in terms of the Full Width at Half Maximum (FWHM) of a Gaussian intensity profile fitted to the source; while the angular scales (e.g. $\theta_{\rm r}$, $\theta_{\rm s}$ etc) used in the scintillation literature are the standard deviation of the fitted Gaussian \citep{Narayan+1992}.}. This is approximately a factor of 10 -- 100 smaller than the implied angular size for rapidly scintillating sources such as PKS\,B1257$-$326 and J1819+3845 \citep{Bignall et al.+2006,Dennett-Thorpe de Bruyn+2003}. Interestingly, in 1999 \citet{Kedziora-Chudczer et al.+2001a} found that the scintillation timescale changed by a factor of 3 -- 4 between 8.6 and 4.8\,GHz, exactly as expected for a point source (Equation~\ref{time}). This implies a source angular size less than $12$\,$\mu$as, and suggests significant source evolution between the May 1999 epoch and our observations in January 2005. The long-term COSMIC data \citep{Carter et al.+2009} indeed show a significant monotonic rise in flux density for the source, beginning some 5 -- 6 months before our epoch of observations. This is suggestive of a new outburst, with a $\mu$as jet component expected to move away from the compact core over time. At a redshift of $1.048$, the angular diameter distance is 1.65~Gpc, and a source size of 20\,$\mu$as corresponds to a physical size of 0.15~pc. Expansion at $0.6c$ is required for such a source evolution. Thus, it is a plausible scenario.

The distance to the phase screen is
\begin{equation}
D_{\rm screen}=\frac{c}{2\pi{\theta_{\rm r0}}^2 \nu_0}
\label{dist}
\end{equation}

Using the \citet{Cordes Lazio+2001} model, we get $D_{\rm screen}=0.86$\,kpc. This is a factor of 100 more distant than the screen distances typically inferred from time delay and annual cycle fitting for the rapidly scintillating sources. For unquenched scintillation, the transverse speed of the phase screen relative to the line of sight to the source is related to this distance, the angular scale of the source $\theta_{\rm s}$ and the corresponding characteristic timescale $t_{\rm char}$ by

\begin{equation}
V_{\rm screen}= D_{\rm screen}(\theta_{\rm s}/t_{\rm char})= D_{\rm screen}(\theta_{\rm r0}/ t_{\rm r0})
\label{speed}
\end{equation}

As discussed above, we observe unquenched scintillation at 4.9\,GHz ($\theta_{\rm s}(4.9)=25$\,$\mu$as) with a timescale $t_{\rm char}(4.9)=1.2$~days. This yields a transverse speed of the phase screen is $V_{\rm screen}=30$~km\,s$^{-1}$. As a final sanity check, we also know that the scintillation is quenched at $6.7$~GHz, and so the screen velocity must exceed $D_{\rm screen} \theta_{\rm s}(6.7)/t_{\rm char}(6.7)=18$~km\,s$^{-1}$.

We can also use the source size inferred from scintillation to estimate the brightness temperature of the most compact emission.  Using the approach outlined in \citet{Walker+1998} we find that at a frequency of 4.9 GHz the implied angular size of the scintillating component (41\,$\mu$as) corresponds to a brightness temperature of $6.2 \times 10^{12}$ K, a factor of 2500 less than that implied by an intrinsic interpretation of the variability.  At a redshift of 1.048 a modest \citep{Hovatta et al.+2009} Doppler factor of 3.8 is required to reduce the brightness temperature in the source frame to beneath the inverse Compton limit.  It is also worth noting that the inferred angular size of the region containing the varying flux density if we assume the variations are intrinsic is $<$ 1 $\mu$as, and hence we would expect to see scintillation from this source in addition to any intrinsic changes.

\section{Conclusions}
Scintillation has been conclusively identified as the cause of short timescale variability at radio frequencies in a number of sources. Some of these (e.g. PKS\,B1257$-$326 and J1819+3845) show variability with characteristic timescales as short as a few hours, but sources with longer timescales such as PKS\,B1519$-$273 and PKS\,B1622$-$253 \citep[e.g.][]{Carter et al.+2009} are much more common \citep{Lovell et al.+2008}. PKS\,B1144$-$379 is a member of a small class of extreme scintillators which are strong ($>$ 1~Jy), have high modulation index ($m > 0.05$) and vary on timescales longer than a day.  Such sources can be monitored with single-dish radio telescopes \citep[as has been demonstrated by][]{Carter et al.+2009}. That short-timescale scintillators are much rarer than longer-timescale ones implies that suitable nearby screens (within a few 10's of pc) are relatively rare, since if such a screen is present the requirements for a background source to scintillate should be met by a large fraction of AGN. We thus expect a large number of more distant screens (the Cordes \& Lazio (2001) model predicts screen distances of 0.5 -- 3~kpc), but these will produced high modulation index scintillation (i.e. not heavily quenched) only for the most compact AGN. So when seeking the most compact AGN either to test physical models, or as targets for space Very Long Baseline Interferometry missions, those sources with high modulation indices and the longest scintillation timescales would appear to be the best targets.

The 1.2 day characteristic timescale (corresponding to a peak-to-peak period of 7.7 days) and high modulation index observed in PKS\,B1144$-$379 make it one of the most extreme bright scintillators identified to date. However, there are likely to be other sources with still longer timescales yet to be identified. The high-cadence, long timescale monitoring of AGN being undertaken by the COSMIC project is ideally suited to identifying such sources, and other rare but brief phenomena such as extreme-scattering events \citep[e.g.][]{Senkbeil et al.+2008}.  Our detection of a distant ($\sim 0.86$ kpc, according to the Cordes \& Lazio model) screen was only possible due to our 15 day monitoring campaign. It is entirely possible that such screens are ubiquitous, and simply not picked up in the short term (2 to 3 days) intensive, or longer term low-cadence observing campaigns which are more commonly undertaken.

\acknowledgements{}
We thank the anonymous referee for a number of thoughtful comments that helped improve the paper significantly. RT is grateful to the University of Tasmania for a Dean's Summer Research Scholarship. SS and JM thank the ARC for Super Science Fellowships. The National Radio Astronomy Observatory is a facility of the National Science Foundation operated under cooperative agreement by Associated Universities, Inc. This research has made use of the NASA/IPAC Extragalactic Database (NED) which is operated by the Jet Propulsion Laboratory, California Institute of Technology, under contract with the National Aeronautics and Space Administration.

\end{document}